# Impact of plasmonic modes on the formation of self-organised nano-patterns in thin films

Panagiotis Lingos[a], George Perrakis[a], Odysseas Tsilipakos[b,a,*], George D. Tsibidis[a,c,**], Emmanuel Stratakis[a,d,]

[a]*Institute of Electronic Structure and Laser Foundation for Research and Technology-Hellas Crete Greece*
[b]*Theoretical and Physical Chemistry Institute National Hellenic Research Foundation GR-11635 Athens Greece*
[c]*Department of Materials Science Technology University of Crete 71003 Heraklion Crete Greece*
[d]*Department of Physics University of Crete 71003 Heraklion Crete Greece*

**Abstract**

Formation of nanoscale laser-induced periodic surface structures on thin metal films (of the size of the optical penetration depth) is a yet unexplored area that is expected to open new routes for laser patterning and a wealth of exciting applications in optics, photonics, and sensing. In contrast to the common belief that excitation of Surface Plasmon Polaritons (SPPs) on the air/metal interface plays the dominant role in the features of the induced topographies, in this work, we demonstrate that the excitation of coupled SPPs in both air/metal and metal/substrate interfaces, along with other parameters such as the thickness of the material, the photon energy, and the substrate refractive index, dictate the spatial modulation of the absorbed energy. A detailed theoretical analysis of the excited plasmonic waves and a multiscale modelling of laser-induced physical phenomena manifests that depending on the laser conditions and thickness of the irradiated solid, topographies with periodic features of diverse sizes (ranging from $\lambda_L/3$ to $\lambda_L$, where $\lambda_L$ stands for the laser wavelength) and different orientation can be realized. The capability to control and tune the characteristics of the produced structures on thin films is expected to enable novel surface patterning

[*]otsilipakos@eie.gr
[**]tsibidis@iesl.forth.gr
stratak@iesl.forth.gr

approaches.

## 1. Introduction

Over the past decades, a plethora of nano/micro structured topographies have been fabricated with femtosecond pulsed lasers by exploiting the wealth of possibilities that laser technology can offer. One particular category is the laser induced periodic induced surface structures (LIPSS), ranging from Low Spatial Frequency (LSFL) to High Spatial Frequency (HSFL) LIPSS (see [1] and references therein) with periodicities $\lambda_L/2 < \Lambda < \lambda_L$ and $\Lambda > \lambda_L/2$, respectively, where $\lambda_L$ stands for the illuminating laser wavelength. LIPSS represent a precise and single-step method to fabricate highly-ordered, multi-directional and complex surface structures. The evident capabilities of the technique to produce self-assembled patterns with impressive functionalities similar to those exhibited in nature has enabled LIPSS technology to receive significant attention due to its important technological applications such as in solar cells, biosensing, advanced optics, wettability, anti-fouling, friction, self-cleaning, tissue engineering, anti-bacterial, adhesion [1–5].

While a control of the modulation of the laser parameters is required to generate an abundance of LIPSS topographies, it has been postulated that the pattern periodicity is ascribed to the interaction of the electromagnetic radiation with the irradiated solid. The widely accepted driving mechanism for LIPSS formation is the interference between the incident light and the excited surface waves such as Surface Plasmon Polaritons (SPP) [1, 3, 6–10] and/or evanescent cylindrical waves that are generated by scattering of light on nano-defects and inhomogeneities [1, 11, 12] on the surface. A detailed investigation of the corresponding radiative and non-radiative fields at the nano-scale (in the vicinity of the defects) has been performed for various materials [1, 12–16].

Nevertheless, the current research on laser-based patterning is predominantly focused on the analysis of the mechanisms leading to LIPSS formation on *bulk* materials. On the other hand, due to the increasing interest in patterning of thin solid films (of sizes comparable to the optical penetration depth) for targeted applications related to optics, healthcare, sensing, environment, energy [17–21], further exploration of possibilities of fabricating

LIPSS patterns on thin materials is imperative. Given the significant impact of the electromagnetic phenomena on the LIPSS fabrication, special attentionis required to analyse the role of near/far field waves that are generated on thin films that comprises a number of defects. In metallic thin films coupled SPPs on the air/metal and metal/substrate interfaces are supported [22–24]. Previous experimental studies demonstrate that irradiation of Au thin films with femtosecond pulses lead to the formation of distinct topographies with thickness dependent periodicities [25]; by contrast, periodic patterns in bulk Au [26] are not significantly pronounced as the electron diffusion cancels the initial spatial modulation in the absorbed energy profile [16]. Furthermore, it is known the dielectric parameters that influence the features of the electromagnetic modes are thickness dependent with an additional impact on the energy absorption from the material [27, 28]. Hence, to provide a detailed analysis of LIPSS formation on thin metals, along with the dominant contribution of the electromagnetic surface waves that are excited on rough thin metal films, it is important to incorporate the impact of thermal and hydrothermal phenomena that lead to pattern formation [6, 13, 16, 29]; such an investigation is aimed to reveal whether distinct periodic structures are formed.

To address the above challenges, in this work, we provide a detailed analysis of the electromagnetic modes that are excited after irradiation of thin gold (Au) films of various thicknesses with femtosecond pulses. Corrugation of the surface that leads to scattering of light is emulated by the inclusion of either a single or many, randomly distributed nanobumps. The role of the thickness, the laser photon energies, the substrate on the excited electromagnetic modes on the surface/interface and the impact on the induced patterns will be highlighted; furthermore, the production of complex (hierarchical) intensity profiles will also be discussed. To project the effect of the electromagnetic phenomena and the absorbed energy spatial distribution on the surface modification, results of the self-organisation mechanism and multiscale (i.e. on various temporal regimes) processes that lead eventually to LIPSS topographies will be presented. It is emphasised that without loss of generality, the current study is focused on subablation conditions.

## 2. Results

### 2.1. Surface plasmon polaritons in thin films

The formation of laser-induced nanoscale periodicities in bulk metals is mainly attributed to the excitation of SPPs on the metal-dielectric inter- face. The interference between the surface waves propagating along different directions on the interface, as well as with the incident field, dictates the resulting field distribution. For the formation of SPPs, two basic condi- tions must be fulfilled. The first one concerns the material properties and requires that the metal permittivity is negative (i.e., operating below the plasma frequency) and, moreover, larger in absolute value than the dielectric(e.g., glass) substrate Re($\epsilon_m$)| > $\epsilon_g$. The second one concerns the abilityto excite SPPs, since there is a momentum mismatch with incident photons.Among other approaches [30, 31], momentum matching can be achieved bysurface roughness and other specific corrugations (e.g., bumps, gratings, etc.) [32, 33].

In metallic thin films, the electromagnetic landscape is modified by the presence of two metal-dielectric interfaces (e.g., with a glass substrate and an air superstrate, respectively) and the ability to support coupled SPPs of even or odd symmetry when the film thickness (*d*) is thin enough compared to the penetration depth. Importantly, the properties of the supported SPPs (prop- agation constant *β* or, equivalently, parallel wavevector) can be controlled by the film thickness. This can be clearly seen in the dispersion relation of *flat* thin films in an asymmetric dielectric environment [34]

$$\exp(-2k_m d) = \frac{k_m/\epsilon_m + k_a/\epsilon_a}{k_m/\epsilon_m - k_a/\epsilon_a} \cdot \frac{k_m/\epsilon_m + k_g/\epsilon_g}{k_m/\epsilon_m - k_g/\epsilon_g}, \qquad (1)$$

where

$$k_j = \xi_j \sqrt{\beta^2 - \epsilon_j k_0^2} \qquad (2)$$

In Eq. (1), *j* = *a, m* or *g* denotes the three media (*a* for air, *g* for glass, *m* for metal), *d* is the thickness of the film, $\xi_j$ = ±1 depending on the mode type, *β* is the propagation constant of the SPP, and $k_0$ = 2π/$\lambda_L$ is the free-space wavenumber at the laser wavelength. Eq. (1) can be solved numerically to find the supported SPP solutions; each solution is associated with an SPP wavelength given by $\Lambda_{SPP}$ = 2π/Re(*β*). For the two bound modes [i.e., using $\xi_a$ = $\xi_g$ = 1 in Eq. (2)], the dependence of the SPP wavelength on film thickness (geometric dispersion) is depicted in Fig. 1. In that figure, the inset illustrates a schematic of all layers considered in the electromagnetic simulations (air/Au/SiO$_2$). It has been calculated at the wavelength of $\lambda_L$ =1026 nm using the data for gold from Ref. [35] and $\epsilon_a$ = 1, $\epsilon_g$ = 2.1. We have used Au as a characteristic metal used in broad range of applications, including various plasmonic waveguides [36, 37], since it exhibits relatively low resistive losses and moreover does not oxidize. For large thicknesses, the two branches asymptotically approach the uncoupled single-interface SPPs on the two interfaces. Interestingly, as the thickness decreases below 30 nm, the lower SPP (SPP″) periodicity decreases abruptly from 687 nm to 395 nm at *d* = 5 nm. The upper branch (SPP′) experiences a smaller variation and it is towards larger periods. The solid lines in Fig. 1 have been calcu- lated assuming that the optical properties of Au are independent of the film thickness. With orange markers we include results using thickness depen- dent gold permittivity, $\epsilon_m(d)$, taken from experimental data in the literature (see Supplemental Material, Section S1). It can be seen that the predicted periodicities almost coincide.

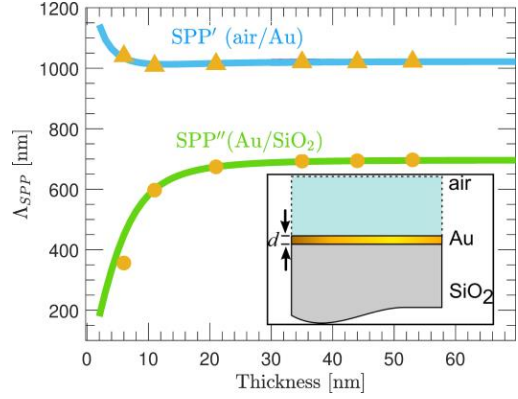

Figure 1: Wavelength of the two bound SPPs on an air/Au/SiO$_2$ thin film at $\lambda_L$ =1026 nm. The upper (SPP′) and lower (SPP″) branch correspond to plasmons of even and odd parity, respectively. The branches asymptotically approach the uncoupled single-interface SPPs on the air/Au and Au/SiO$_2$ interface, respectively. The solid lines depict the pure dependence on the Au film thickness by using the thickness independent Au permittivity from Ref. [35]. With markers we plot results with thickness dependent permittivity data for Au (see Supplemental Material, Section S1). Inset shows schematic of layers considered in the electromagnetic simulations.

## 2.2. Optical response and interference patterns from scattering by a single nano-bump

To explore the features of the surface waves that are excited due to inhomogeneities on flat surfaces, we first study the case of a single scatterer by numerically calculating the field distribution for various film thicknesses. In all cases, the scatterer is a nano-sized hemispherical bump of radius $r = 50$ nm residing on the top air/Au interface. A bump scatterer is selected instead of a hole, since in ultra-thin films the hole would lead to direct coupling with the bottom interface. Smaller bumps would result in similar distributions but with a less efficient excitation of the SPPs (less absorption). Larger nanobumps would lead to multipolar localized modes. To calculate the electromagnetic (EM) field distribution, we solve the integral form of the Maxwell equations [38, 39] considering thickness dependent optical properties of the metallic films (see Supplementary Materials, Tables S1 and S2). The laser radiation is normally incident and linearly-polarized along the $x-$axis and the wavelength is $\lambda_L = 1026$ nm or $\lambda_L = 513$ nm. A uniform distribution of the incident EM field has been considered, which would physically correspond to a collimated beam; this maximizes the interaction between the incident pulse and the nanorough gold surface. Note that the resulting LIPSS period is not expected to be strongly affected compared to a focused laser beam[40]. For additional simulation details see the Supplementary Material (Section S2). We use the intensity (power density), $L E^2$, to capture the spatial profile of the energy absorption (Joule losses) in the metallic film. In order to focus on the effect of the nanobumps, the intensity maps are shown normalized to the intensity on a flat Au surface, $I_S$. The quantity that is plotted is $(I - I_S)/I_S$. Maps are shown which are normalized to the intensity on a flat Au surface, $I_S$.

In Fig. 2 we plot the intensity enhancement (compared to a flat metal surface) in the $xy$ plane just below the air/Au interface for film thicknesses $d = 6, 11, 21, 35, 44$ nm and thick (bulk) material. The intensity distribution is indicative of the energy absorbed by the material due to resistive loss, which, in effect, acts as a heat source and can even lead to material melting and movement. We have chosen these specific thicknesses since experimental data regarding the thickness dependent permittivity are available in the literature [41–43] (see Supplementary Material for more details). In all cases, far from the nano-bump we observe periodic patterns of field minima and maxima alternating *along* the polarization direction ($x$ axis), characteristic of excited SPP surface waves, which are necessarily TM polarized. The near field ($r < \lambda_L/2\pi$) shows strong enhancement and exhibits a more complicated distribution, extending to the perpendicular (*y* axis) direction as well (especially panels (a)-(c) in Fig. 2).

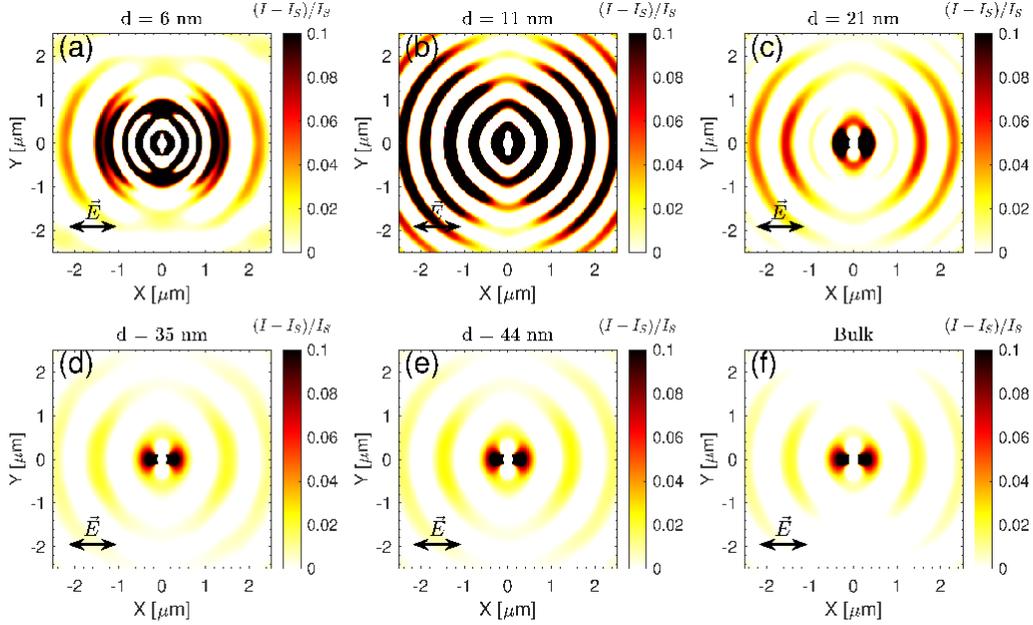

Figure 2: Scattering by a single nano-bump for different metal film thicknesses (air/Au/SiO$_2$ structure). Intensity distribution in the *xy* plane just below the top air/Au interface for $\lambda_L$ = 1026 nm. The optical properties of Au are thicknesses-dependent ([41– 43], see the Supplemental Material). The corresponding spatial Fourier spectra can be found in the Supplemental Material. Double-headed arrow indicates the orientation of the laser beam polarization.

For thicknesses *d* = 35 nm [Fig. 2(d)] and *d* = 44 nm [Fig. 2(e)], the distribution resembles that of the bulk case [Fig. 2(f)]. Further analysis with the help of a spatial Fourier transform (FT) reveals that both $\Lambda_{SPP'}$ and $\Lambda_{SPP''}$ periodicities are observed, with the $\Lambda_{SPP'}$ (upper branch in Fig. 1, long wavelength) being significantly more prominent, see Supplementary Material (Section S3). This indicates that for such thicknesses coupling with the bottom (Au/SiO$_2$ interface) is weak, as expected. On the other hand, as the film thickness decreases the periodicities reflected on the field distribution begin to change significantly as can be clearly seen in Fig. 2(a)-(c). Computing the FT (see Supplemental Material), we find that the predominant periodicities are approximately 674, 600, and 365 nm for *d* =21, 11, and 6 nm, respectively. These now correspond to the SPP″ branch (cf. Fig. 1)and are characterized by the sharp decrease of the SPP wavelength as thefilm thickness decreases. Periodicities that correspond to the other, SPP′, branch can be also seen in the Fourier spectra, but they have become weaker for these small film thicknesses. In addition, for small thicknesses ($d$ =6, 11, 21 nm) the reciprocal space reveals frequency content along the $y$ axisas well, i.e., perpendicular to the polarization (see Supplemental Material). This is connected with the near-field scattering from the nanobump, as canbe verified by looking at the direct space [Fig. 2(a)-(c)].

In order to understand the impact of (i) the film thickness itself and (ii) the thickness dependent gold permittivity on the field distributions of Fig. 2, we have repeated the simulations adopting a constant permittivity $\epsilon_m$ = 44.472 + $i$3.2056 throughout, taken from Ref. [35] for 1026 nm. The corresponding distributions in direct and reciprocal spaces can be found in the Supplementary Material (Section S3). It can be seen that the main features remain qualitatively the same as in the thickness dependent permittivity case. The most notable difference is for $d$ = 6 nm. In this case, one can see that the SPP″ wavelength remains very prominent. This is because the imaginary part of the permittivity remains low (it increases in the thickness dependent data) leading to a long propagation length for the SPP″ wave.

The above study has been, also, conducted at a different operating wavelength $\lambda_L$ = 513 nm and the obtained field distributions are compiled in Fig. 3. At a such short wavelength which approaches the plasma frequency, the metal can become quite transparent and the condition for supporting SPPs ( $\text{Re}(\epsilon_m) > \epsilon_g$) may cease to hold. As a result, the presence of 'ra- diative', spherical scattering by the nanobump (not in the form of excited surface waves) is anticipated to be stronger. Indeed, in Fig. 3 one can see strong periodic modulation along the $y$ (perpendicular) axis which is not limited to the near field of the scatterer, indicating the presence of sphericalscattered waves (SPPs would favor the $x$ direction). This is also reflected in the Fourier spectra, see the Supplementary Material (Section S3). Moreover, it is interesting to note that the exact structure of the near field distributioncan be elegantly controlled by the film thickness. This can become importantin the case of multiple-bump scattering (see Section 2.3) and can affect the evolving surface topography upon multi-pulse irradiation.

The results presented already in this Section demonstrate that the extra degree of freedom introduced by the variable thickness of the film can provide new opportunities compared to bulk metals. The additional advantage will become more evident in the following Sections (2.3 and 2.4), where further complexity will be introduced by increasing the corrugation of the film surface.

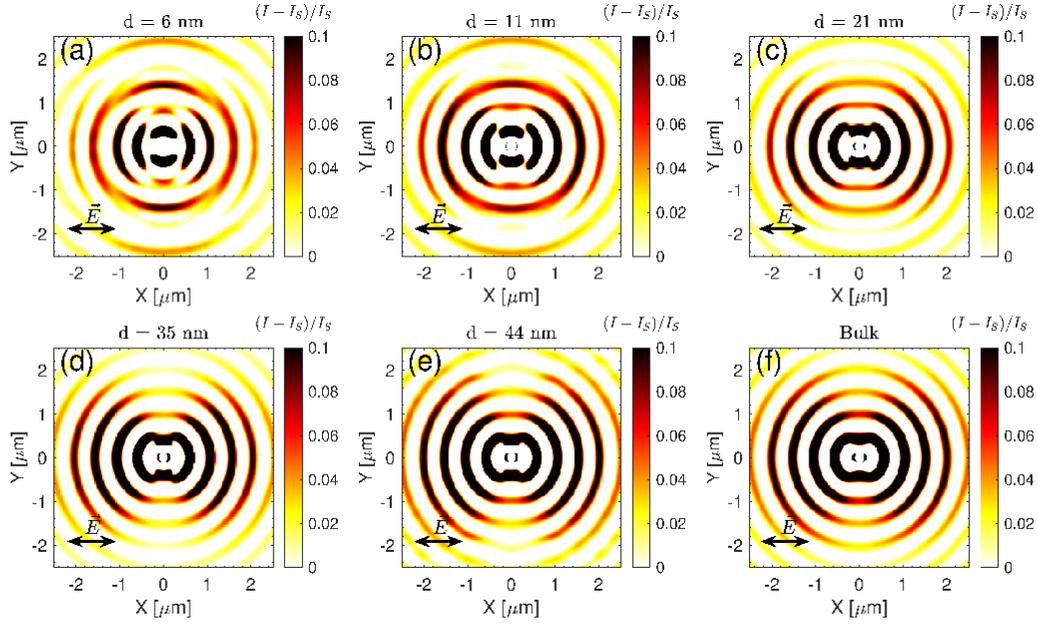

Figure 3: Scattering by a single nano-bump for different metal film thicknesses (air/Au/SiO$_2$ structure). Intensity distribution in the *xy* plane just below the top air/Au interface for $\lambda_L$ = 513 nm. The optical properties of Au are thicknesses-dependent and taken from experimental data in the literature [41–43], see the Supplemental Material. The corresponding spatial Fourier spectra can be found in the Supplemental Material. Double-headed arrow indicates the orientation of the laser beam polarization.

### 2.3. Optical response and interference patterns from scattering by randomly distributed nano-bumps

In the previous section, we investigated the interference patterns arising from the interaction of the incident beam with a single nano-bump resid- ing on the (top) air/Au interface and focused on the impact of the metallic film thickness. Since a realistic (corrugated) metallic surface contains a high concentration of randomly distributed surface defects (arising from surface roughness and/or induced structuring due to repetitive irradiation), to elucidate the role of the collective response of such nano-defects we perform simulations for a given, randomly-distributed nano-bump pattern with specific concentration and various film thicknesses and we analyze the resulting periodic features by studying the corresponding spatial FT. In the following, we consider a bump concentration $C = N\pi r^2/S \approx 0.3\%$ where $N$ is the number of nano-bumps with radius $r$ = 50 nm and the laser affected area $S = 8 \times 8$ $\mu$m$^2$. The above concentration corresponds to an average spacing between nano-bumps $D = \sqrt{\pi r^2/C} \approx 1.6$ $\mu$m. For such a long average spacing distance $D > \lambda_L$, the nano-bumps behave as uncoupled dipoles that interact via their far field scattering. However, in some regions the neighboring bumps are spaced by distances smaller than $\lambda_L/2$, in which case their near fields do also overlap. In Fig. 4 we depict the intensity distribution for various film thicknesses $d$ = 6, 11, 21, 35, 44 nm; results for bulk Au are also included for comparison purposes. The corresponding Fourier spectra can be found in the Supplemental Material (Section S4).

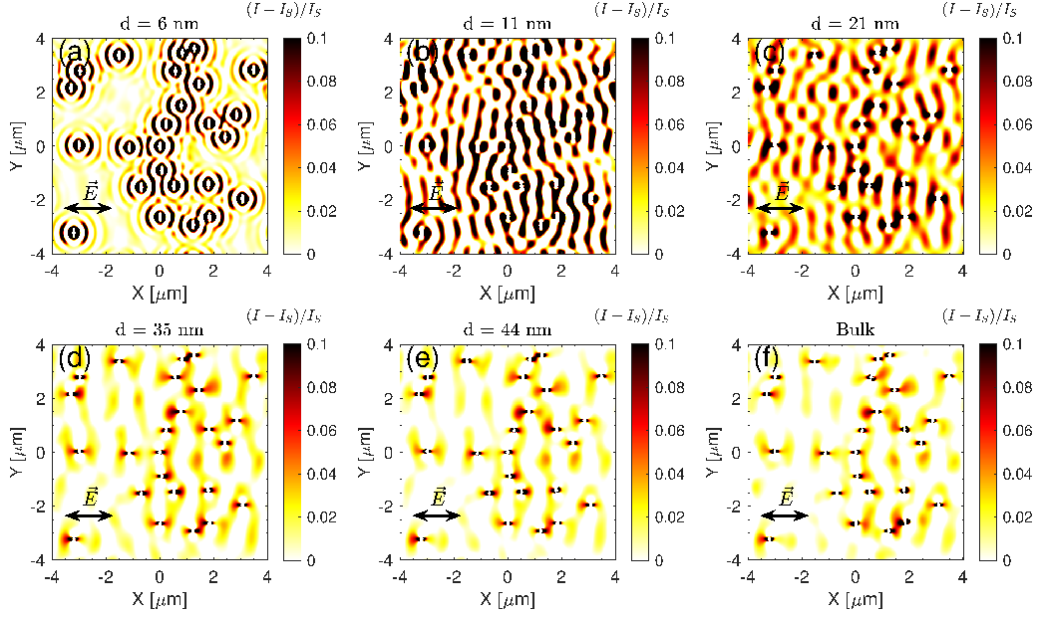

Figure 4: Scattering by randomly distributed nano-defects of radius $r$ = 50 nm and average spacing $l \approx$ 1.6 $\mu$m for different metal film thicknesses (air/Au/SiO$_2$ structure). Intensity distribution in the *xy* plane just below the top air/Au interface for $\lambda_L$ = 1026 nm. The optical properties of Au are thicknesses-dependent and taken from experimental data in the literature [41–43], see the Supplemental Material. The corresponding spatial Fourier spectra can be found in the Supplemental Material. Double-headed arrow indicates the orientation of the laser beam polarization.

The results depicted in Fig. 4 show that periodic features develop along the $x$-axis (the ripples are perpendicular to $E$-field polarization), as anticipated due to the excitation of TM-polarized SPP surface waves. Looking at the Fourier spectra we notice that for $d \geq 35$ nm it is the SPP′ (long period) which dominates, similarly to the single bump (see Section 2.2). In contrast, for smaller thicknesses coupling with the bottom interface becomes stronger and the SPP″ wave becomes prominent. This behaviour can be verified by the Fourier spectra (Supplementary Material), where the major frequency content is found near $k_x/k_0 \approx \lambda_L/\Lambda_{SPP″}$. The thickness $d = 21$ nm is the 'transitioning' thickness, with the two SPPs being roughly equally represented in the Fourier spectrum. Importantly, for $d = 11$ nm [Fig. 4(b)], the SPP″ periodicity and the mean bump separation are commensurate, giving rise to a uniformly rippled distribution. In addition, the near fields of individual scatterers and the interference between them depending on the relative positions of neighboring nanobumps are reflected in a diffused ensemble of even longer wavevectors, as is characteristic of roughness type ($r$-type) HSFLs. Interestingly, for a very thin film such as $d = 6$ nm [Fig. 4(a)], we observe the scattering of spherical 'radiative' waves; the corresponding Fourier spectrum encompasses directions towards the $y$ axis.

A key question is the level at which a change in the illuminating (laser) wavelength affects the electromagnetic response of the material. Thus, an analysis is also performed for $\lambda_L = 513$ nm to identify any difference in the electromagnetic fingerprint on the irradiated solid. The average nanobump spacing is modified to 0.676 $\mu$m ($C \approx 1.7\%$), in order to lead to similar electrical spacing (distance in units of free-space wavelength) as in the 1026-nm case. Simulation results for $d = 6, 11, 21, 35, 44$ nm are summarized in Fig. 5. In this case, the interference between the near fields of each nanobump appears to be very important. As we have observed in Fig. 3, the exact structure of the near fields can be tuned by the film thickness and extends in both the $x$ and $y$ axes. This gives rise to a very interesting distribution for $d = 11$ nm [Fig. 5(b)], where the periodic features develop primarily along the $y$ axis, instead of the $x$ axis. The development of ripples which are parallel to the laser polarization is a feature commonly observed in dielectric materials. However, we find that metallic films of thickness shorter than the optical penetration depth can also emulate such patterns. We anticipate that for a given bump concentration we can always find a suitable/optimal thickness for achieving such a peculiar field distribution. Given that the bump concentration is related to the corrugation of the irradiated surface, which, in principle, is generated in specific laser conditions (i.e., fluence, number of pulses, etc.), our results manifest the capacity to control and tune the orientation and size of the produced structures according to thethickness of the metal. This is a characteristic example of new opportunitiesfor patterning in thin metal films through varying the thickness, in contrast to bulk samples (see Section 2.5).

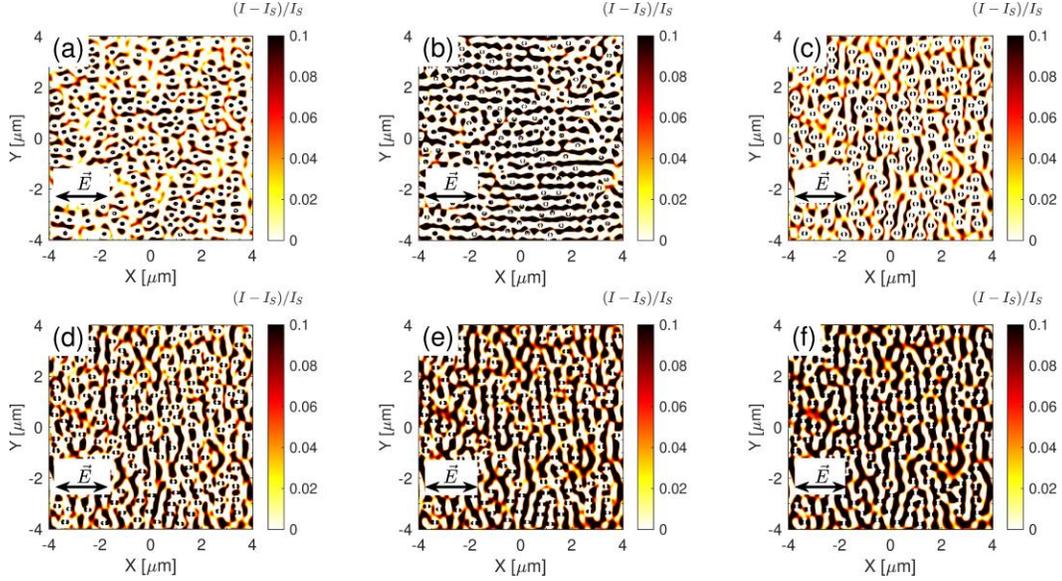

Figure 5: Scattering by randomly distributed nano-defects of radius $r = 50$ nm and average spacing $l \approx 0.676$ $\mu$m for different metal film thicknesses (air/Au/SiO$_2$ structure). Intensity distribution in the *xy* plane just below the top air/Au interface for $\lambda_L = 513$ nm. The optical properties of Au are thicknesses-dependent and taken from experimental data in the literature [41–43], see the Supplemental Material. The corresponding spatial Fourierspectra can be found in the Supplemental Material. Double-headed arrow indicates the orientation of the laser beam polarization.

## *2.4. Hybrid patterns*

A particularly attractive and challenging goal in the field of materials processing is the formation of hybrid topographies exhibiting different topographical hierarchies upon femtosecond laser irradiation [44]. A variety of techniques and approaches have been proposed for the realization of morphologies exhibiting a range of spatial frequencies, since their multipronged nature can offer distinct advantages over single spatial frequency ones for a variety of applications and future concepts such as optical and thermal systems, water repellent and antimicrobial surfaces [45, 46]. We have already shown that the high versatility of the thin-film platform can result in energy deposition with hybrid periodic features absent in bulk materials and without requiring elaborate designs or techniques. For instance, the two supported SPPs can give rise to the simultaneous presence of two periodicities on the laser processed area, while their superposition and relative contribution can be controlled by a proper variation of the film thickness. As such, we can achieve hybrid topographies in a manner similar to direct laser interference patterning (DLIP), but without the employment of overlapping two or more laser beams for creating interference patterns or combining several different techniques to indirectly treat the material's surface [47]. This synergy of the two SPPs becomes even more interesting when their corresponding periods are quite different. The requirements for achieving such conditions by tuning the film thickness and the bump concentration are discussed below. Furthermore, the impact of the differences between the refractive indices of air and the substrate on the field distribution is also presented.

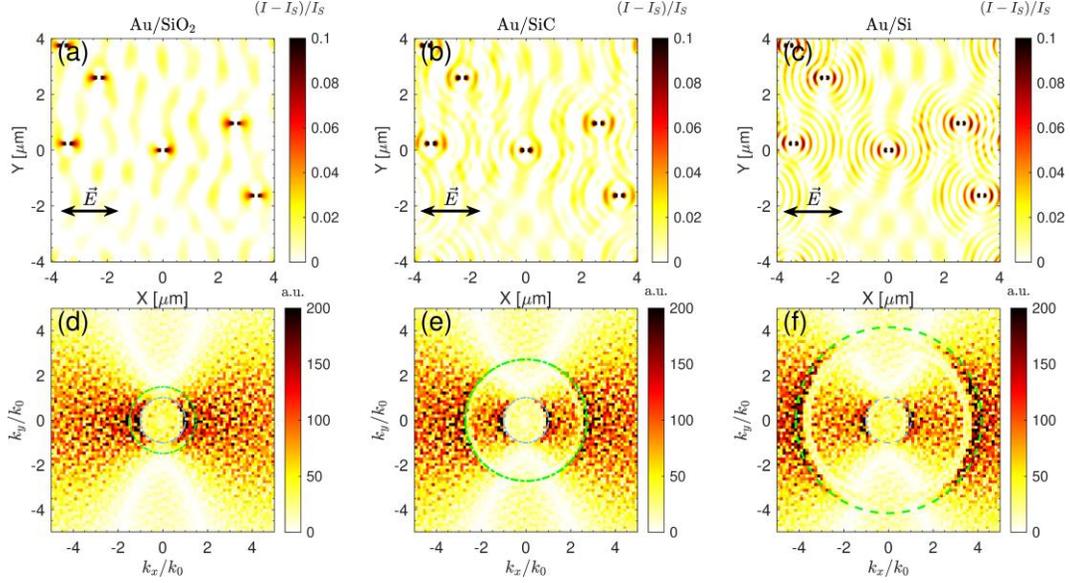

Figure 6: Scattering by randomly distributed nano-defects of radius $r$ = 50 nm and average spacing $l \approx$ 1.6 $\mu$m different refractive indices of the substrate: (a) silicon dioxide ($n_{sub}$ = 1.45), (b) silicon carbide ($n_{sub}$ = 2.5), and (c) silicon ($n_{sub}$ = 3.5). The intensity distribution in the *xy* plane is plotted for a film thickness $d$ = 35 nm and laser illumination at $\lambda_L$ = 1026 nm. As the refractive index difference between superstrate and substrate increases, two quite distinct periodicities are imprinted on the field distribution and Fourier spectrum. Double-headed arrow indicates the orientation of the laser beam polarization.

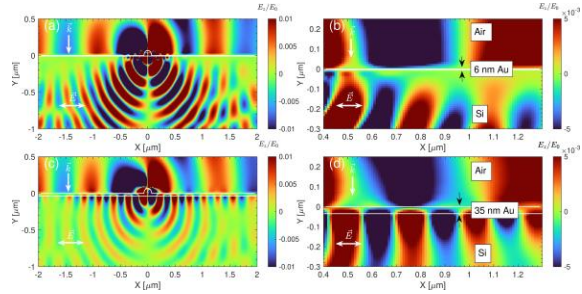

Figure 7: Electric field (*z*-component) distribution in an *xz* plane of the air/Au film/dielectric system for (a) *d* = 6 nm (an enlarged area is illustrated in (b)), (c) *d* = 35 nm

Predicting a priori the effect of the film thickness on the formation of a hybrid topography is not straightforward, since there are several interference phenomena involved. The ability to couple two SPPs via the nanobump scatterers is of particular importance, which depends on the disparity between the wavevectors of SPPs and free-space photons and, thus, on the SPP dispersion and the operating wavelength. Moreover, other factors such as the defect concentration and the particular metal/dielectric material properties give rise to an even more complex electromagnetic behavior. For instance, too small defect concentrations may hinder the formation of a uniform hybrid topography, due to the short-lived nature of the tightly-confined, short-period SPP. On the other hand, too large defect concentrations may severely quench the scattering from individual nanobumps. All these effects are mostly unexplored and need careful examination so as to conclude on their impact in realizing combined hierarchies with a plasmonic film and, subsequently, be able to control the resulting distribution.

In order to study these issues, we have varied several parameters (substrate refractive index, $n_{sub}$, defect concentration, laser wavelength), in order to identify different distribution regimes and optimum conditions. In what follows, we show results for different substrates and a wavelength $\lambda_L$ = 1026 nm. More specifically, the results compiled in Fig. 6 discuss a 35-nm film, which, in our case, produces the clearest hybrid response, for three types of substrates: fused silica ($n_{sub}$ = 1.45), silicon carbide (SiC) ($n_{sub}$ = 2.5) and silicon ($n_{sub}$ = 3.5). Furthermore, Fig. 7 presents a comparison with a different film thickness, showing that for *d* = 35 nm there is a simultane-ous presence of two quite different periodicities on the top interface, which, moreover, survive for long propagation lengths. Returning to Fig. 6, the refractive index asymmetry between superstrate (air in this case) and substrate is an important factor in achieving a clear imprinting of the two SPP periodicities on the intensity distribution. This is clearly demonstrated in the Fourier spectra of Fig. 6(d)-(f): for the case of $n_{sub}$ = 3.5 [Fig. 6(f)], the two SPP wavevectors are further apart and nicely distinguished and diffusionin $k$-space is minimized.

*2.5. Formation of self-organised periodic patterns*

As discussed above, the interaction of light with the surface inhomogeneities on a thin film produces electromagnetic interference patterns on the surface which results into a spatially periodic energy deposition. The evaluation of the deposed energy (and absorbed) energy through the calculations performed in the previous sections provide a first step in a multi-physical study towards explaining the formation of self-organised periodic structures. More specifically, it is known that surface patterning with femtosecond pulsed lasers is a multiscale process that comprises physical phenomena such as energy absorption, electron excitation, relaxation, phase transitions and resolidification [6, 14]. Hence, any theoretical framework aiming to correlate the laser parameters/material characteristics with surface modification has to include and couple modules that describe the above mechanisms [6] that occur at different temporal scales. Calculations for the intensity profiles performed in the previous sections provide the input to the energy absorption module and lead to a precise evaluation of the electron excitation levels, determination of material temperature distribution and characteristics of the induced molten volume and finally, deduction of information about the features of the induced LIPSS. The theoretical model that is used in this work to describe the aforementioned coupled physical processes assumes linearly polarized laser beams of $\lambda_L$ = 1026 nm, pulse duration $\tau_p$ = 170 fs and fluences $F$ =0.2 J/cm$^2$ and $F$ =0.7 J/cm$^2$ for $d$ = 11 nm and bulk materials, respectively (details of the multiscale model are provided in [6, 14, 16] and references therein).The fluence values were appropriately chosen to produce a sufficient molten volume in Au for $d$ = 11 nm and bulk Au. In a recent study, simulation results for the optical properties and damage thresholds were obtained for $\lambda_L$ = 1026 nm that indicate that the aforementioned values are sufficient for the production of a thin layer of molten material [28]. Dynamics of the molten volume and resolidification mechanisms will eventually lead to the production of periodic topographies. It is noted that no ablation con-ditions were considered in this study. Although the theoretical frameworkdescribed above can be also applied in conditions were mass removal is as- sumed, the aim of the current investigation was focused on the illustration of the formation of periodic patterns on thin films by simply utilizing the EMand hydrothermal effects. By contrast, fabrication of topographies on the Au layer assuming ablation conditions is expected to be a rather challenging issue especially for very thin films at least from an experimental point of view (to avoid reaching the substrate through the repetitive irradiation).

In Fig. 8(a,b), simulations results of the induced topographies for $d$ =11 nm (Au/SiO$_2$) and bulk are presented. Similar conclusions can be deduced for irradiation at $\lambda_L$ = 513 nm. The predictions demonstrate that for small thicknesses prominent periodic topographies are realized [Fig. 8(b)], in contrast to a substantially less prominent (nearly absent) LIPSS profile for bulk Au [Fig. 8(a)]. The striking difference is attributed to the fact that thin films hinder the diffusion of hot electrons into deeper regions of the irradiated metals in contrast to a significantly easier electron transport for bulk materials. Therefore, in thin films, a large number of highly energetic electrons remain close to the surface and the gradient of the spatially modulated electron temperature (due to the spatially varying absorbed energy from the electron system) is more enhanced. Upon scattering with the lattice, the excited electrons will transfer part of their energy to the lattice but due to the larger (spatially) temperature gradients, the thermal effects and hydrodynamic phenomena will be sufficiently high to produce pronounced ridges on the pattern in contrast to what happens for bulk. The above argument has been confirmed through the employment of a multiscale theoretical framework [16, 28] while experimental observations for bulk Au illustrate the less distinct LIPSS patterns [26]. A distinctly different periodicity that is dictated from the electromagnetic phenomena is also projected on the topographies for the two cases.

Another aspect that is of paramount importance is, also, the type of excited electromagnetic modes and patterns that are formed following repetitive irradiation of thin films. As mentioned in the introductory section, various types of LIPSS are realized by increasing the number of pulses and the synergetic impact of electromagnetic and hydrodynamic phenomena in the LIPSS periodicity was discussed in laser processing of bulk materials in a previous report [13]. Nevertheless, special attention is required given the fact that there is a spatial variation of the LIPSS height [Fig. 8(a)]. The dis-tribution of distances of the irradiated region from the substrate is expected to yield more complex electromagnetic behaviour. In other words, the topography that is irradiated is equivalent to a metal pattern of (spatially) varying thickness. Thus, the results of Section 2.3 appear to apply in this case, however, a more detailed investigation is required that is outside the scope of the present work.

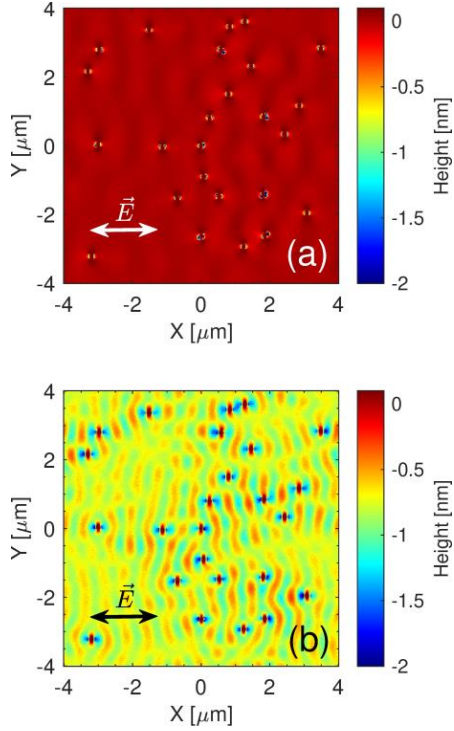

Figure 8: LIPSS patterns for (a) Bulk Au, (b) Au of thickness $d$ = 11 nm. Note the significantly deeper modulation in the thin-film case. Double-headed arrow indicates the orientation of the laser beam polarization.

## 3. Discussion and Conclusions

In this article, we showed the excitation of electromagnetic waves of periodicities of diverse sizes (ranging from $\lambda_L/3$ to $\lambda_L$) and orientation on metallic films of various thicknesses comparable to the optical penetration depth. Simulations results revealed the significant role of the thickness that is also projected on the induced topographies following a multiscale physical model that comprises apart from the energy distribution, the impact of the hydrothermal effects. Special emphasis was drawn on the role of the laser wavelength and the substrate which both demonstrated a significant impact. Theoretical predictions demonstrated that irradiation of a corrugated surface (as emulated with a random distribution of nanobumps) at $\lambda_L$ = 513 nm lead to periodic features *parallel* to the laser polarisation, in contrast to thicker films. Furthermore, for all thicknesses, the interference of the near fields appears to dominate the electromagnetic fingerprint. By contrast, at longer laser wavelengths (i.e., $\lambda_L$ = 1026 nm), the picture does not remain the same: at very small thicknesses spherical radiative waves develop ($d$ = 6 nm), while at $d$ = 11 nm the short-wavelength SPP (SPP″) dominates. For thicker films both SPPs are imprinted ($d$ = 21 nm), before the SPP′ prevails at even larger thicknesses ($d$ > 35 nm). To reveal the significant role of the substrate, the use of three substrates, fused silica, SiC, and silicon showed that as the refractive index asymmetry between air and substrate increases, the possibility for imprinting two distinct periodicities increases. These effects give rise to the formation of more complex patterns.

It is evident that a systematic evaluation of the theoretical simulations and predictions should be complemented with the development and employment of suitable experimental protocols, however, this is beyond the scope of the present work. In regard to the comparison of results from theoretical simulations with experimental data, it is noted that, usually, thin metallic films are considered for adhesion purposes and in the case of Au, thin Cr films are used (normally, of the size of some tens of nm). As a first indication of differences in the presence of a thin Cr layer, the following table ([Table 1]) indicates that a 10-nm thick Cr adhesion layer can have strong impact on the reflectivity and transmission within the $SiO_2$ substrate in some cases. Calculations have been conducted for two wavelengths, 513 nm and 1026 nm while $SiO_2$ is used as a substrate (similar results can be deduced for other substrates). As shown in Table 1, transmittance is lower if Cr is included in the multilayer film. At lower photon energies and for thicknesses larger than the optical penetration depth, the impact of the adhesion layer on the reflectivity ($R$) and transmittance ($T$) appears to be weak, as expected. Thus, in these cases, the optical parameter values considered in the simulations in this study (in the absence of Cr film) can be used to evaluate the energy absorbed in the multilayer structure.

Apart from the electromagnetic response, it is important to evaluate the role of the different thermophysical properties of the materials (in Au/Cr/$SiO_2$ or Au/SiO$_2$) on the thermal fingerprint on the irradiated multilayered solid. In principle, thermal effects are expected to influence the characteristics of the induced pattern. However, the predominant objective of the approach presented in this study was to investigate the interference of the electromagnetic modes on the two layers (Air/Metal and Metal/Dielectric), predict the spatial modulation of the absorbed energy and explore how the coupling of the two modes affect the electromagnetic fingerprint and induced surface pattern. Certainly, the inclusion of the Cr layer will provide a more accurate evaluation of the quantitative characteristics of the absorbed energy and allow a direct comparison with realistic experimental results; nevertheless, for the sake of simplicity, consideration of the role of a Cr layer has been ignored. The latter can be the subject of a future (and more detailed) investigation.

|  | $\lambda_L$=513 nm | | | | $\lambda_L$=1026 nm | | | |
|---|---|---|---|---|---|---|---|---|
|  | Au/Cr/SiO$_2$ | | Au/SiO$_2$ | | Au/Cr/SiO$_2$ | | Au/SiO$_2$ | |
| Thickness | R | T | R | T | R | T | R | T |
| 6 nm | 0.39 | 0.21 | 0.08 | 0.81 | 0.44 | 0.26 | 0.38 | 0.57 |
| 11 nm | 0.43 | 0.18 | 0.14 | 0.68 | 0.63 | 0.16 | 0.65 | 0.29 |
| 21 nm | 0.52 | 0.13 | 0.27 | 0.46 | 0.84 | 0.07 | 0.87 | 0.09 |
| 35 nm | 0.59 | 0.07 | 0.43 | 0.26 | 0.94 | 0.02 | 0.95 | 0.03 |
| 44 nm | 0.61 | 0.05 | 0.5 | 0.17 | 0.96 | 0.01 | 0.96 | 0.01 |
| Bulk | 0.62 | 0 | 0.62 | 0 | 0.98 | 0 | 0.98 | 0 |

Table 1: Reflectivity (*R*) and Transmittance (*T*) of multilayered solids (Au/SiO$_2$) or (Au/Cr/SiO$_2$) for various thicknesses of Au. Calculations for two different wavelengths, $\lambda_L$ = 513 nm and $\lambda_L$ = 1026 nm have been performed.

As different materials were used in Section 2.4 to illustrate the role of the substrate, apart from the variance of the induced electromagnetic modes, it is also important to explore potential distinct surface modification features due to the optical properties of the substrates. More specifically, it is known thagt Si absorbs more than SiO$_2$ which should yield also different laser energy absorption levels in the irradiated metal. This difference is expected to diminish at longer wavelengths ($\lambda_L$ = 1026 nm) As a result, the produced thermal effects are expected to lead to some differences in the morphological features of the pattern (i.e. height). This is an aspect that should be investigated in more detail in a future study. In regard to whether, the laser conditions and substrate-dependent optical properties are expected to yield any modification on the substrate, it is noted that the laser energy was appropriately selected to induce surface modification on the metallic layer. By contrast, the portion of the energy transmitted to the substrate is not sufficient to induce electron excitation or thermal effects that will damage the substrate.

One aspect that should be, also, addressed is the accuracy of the dielectric parameter that was used for the irradiated solid. As pointed out in the description of the theoretical framework, constant values of the optical parameters are considered in this analysis. Due to a temporal variation of the dielectric function [28], a more accurate approach would require the incorporation of the change of the optical parameters as well. To this end, appro- priate pump-probe experiments could be developed for the evaluation of the optical parameter temporal change. Similarly, the employment of Density Functional Theory-based calculations could be performed to derive a tempo-ral variation of the dielectric function [48, 49]. Despite the approximations performed in this work, the prediction results present in a consistent fashion the significant role of the metal thickness in the coupling of the excited elec- tromagnetic modes and the impact on the produced surface topographies.Our results are expected to provide novel routes for optimizing the outcomeof nano-processing of LIPSS structures in thin films for a plethora of potential applications.

## CRediT authorship contribution sttaement

**P.Lingos**: Conceptualisation, Methodology, Investigation, Simulations, Writing-original draft. **G.Perrakis**: Conceptualisation, Methodology, Investigation, Simulations, Writing-original draft. **O.Tsilipakos**: Conceptualisation, Supervision, Methodology, Investigation, Simulations, Funding Acquisition, Writing-original draft. **G.D.Tsibidis**: Conceptualisation, Supervision, Methodology, Investigation, Simulations, Writing-original draft. **E. Stratakis**: Conceptualisation, Funding Acquisition, Writing-review & editing. **PL** and **GP** contributed equally to this work.

## Declaration of Competing Interest

The authors declare the following interests/personal relationships which may be considered as potential competing interests: Emmanuel Stratakis reports financial support was provided by the European Commission.

**Data availability**

Data will be made available on request.

**Funding**

GDT and ES acknowledge support by the European Union's Horizon 2020 research and innovation program through the project BioCombs4Nanofibres (grant agreement No. 862016). GDT and ES acknowledge funding from HELLAS-CH project (MIS 5002735), implemented under the "Action for Strengthening Research and Innovation Infrastructures" funded by the Operational Programme "Competitiveness, Entrepreneurship and Innovation" and co-financed by Greece and the EU (European Regional Development Fund) while PL and GDT acknowledges financial support from COST Action TUMIEE (supported by COST European Cooperation in Science and Technology). OT acknowledges support by the Hellenic Foundation for Research and Innovation (HFRI, http://dx.doi.org/10.13039/501100013209) under the "2nd Call for HFRI Research Projects to Support Post-Doctoral Researcher", Project No. 916 (PHOTOSURF).